\documentclass[conference]{IEEEtran}

 \IEEEoverridecommandlockouts
\usepackage{enumitem}
\usepackage{comment}

\usepackage{cite}

\usepackage{multirow}
\usepackage{amsmath,amssymb,amsfonts}
\usepackage{algorithmic}
\usepackage{graphicx}
\usepackage{textcomp}
\usepackage{stmaryrd}
\usepackage{xcolor}
\usepackage{tabularx,booktabs}

\def\BibTeX{{\rm B\kern-.05em{\sc i\kern-.025em b}\kern-.08em
    T\kern-.1667em\lower.7ex\hbox{E}\kern-.125emX}}

\usepackage{amsmath,amsfonts,bm}

\def\eqref#1{equation~\ref{#1}}

\def\1{\bm{1}}

\DeclareMathAlphabet{\mathsfit}{\encodingdefault}{\sfdefault}{m}{sl}
\SetMathAlphabet{\mathsfit}{bold}{\encodingdefault}{\sfdefault}{bx}{n}

\usepackage{circledsteps}
\pgfkeys{/csteps/inner color=white}
\pgfkeys{/csteps/outer color=none}
\pgfkeys{/csteps/fill color=black}
\usepackage{pifont} 

\definecolor{cadmiumgreen}{rgb}{0.0, 0.42, 0.24}
\usepackage{setspace}
\begin{document}

\title{FPGA-Patch: Mitigating Remote Side-Channel Attacks on \underline{FPGA}s using Dynamic \underline{Patch} Generation}

\author{
 \IEEEauthorblockN{Mahya~Morid~Ahmadi$^\dag$,
        Lilas~Alrahis$^\ddag$,
       Ozgur~Sinanoglu$^\ddag$ and Muhammad~Shafique$^\ddag$}
  	\IEEEauthorblockA{$^\dag$\textit{Technische Universit\"at Wien (TU Wien), Vienna, Austria}\\
  	$^\ddag$\textit{Division of Engineering, New York University Abu Dhabi (NYUAD), Abu Dhabi, United Arab Emirates}\\
  	Email: \ mahya.ahmadi@tuwien.ac.at, \{lma387, ozgursin, muhammad.shafique\}@nyu.edu}\\ \vspace{-20pt}
 }

\maketitle

\begin{abstract}
We propose \textit{FPGA-Patch}, the first-of-its-kind defense that leverages automated program repair concepts to thwart power side-channel attacks on cloud FPGAs. FPGA-Patch generates isofunctional variants of the target hardware by injecting faults and finding transformations that eliminate failure. The obtained variants display different hardware characteristics, ensuring a maximal diversity in power traces once dynamically swapped at run-time. Yet, FPGA-Patch forces the variants to have enough similarity, enabling bitstream compression and minimizing dynamic exchange costs. Considering AES running on AMD/Xilinx FPGA, FPGA-Patch increases the attacker’s effort by three orders of magnitude, while preserving the performance of AES and a minimal area overhead of 14.2\%.

\end{abstract}

\section{Introduction}
\label{introduction}

Field programmable gate arrays (FPGAs) are gaining increasing attention for their accelerated and low-power computations, leading cloud service providers (CSPs) to deploying FPGA platform-as-a-service~\cite{Survey5, Survey4}. However, CSPs consider FPGA virtualization for multi-tenancy, which poses security risks in cloud FPGAs~\cite{Virtualization_2018,cpamap_2020}.

\begin{figure}[!t]
 \centering
 \includegraphics[width=\linewidth]{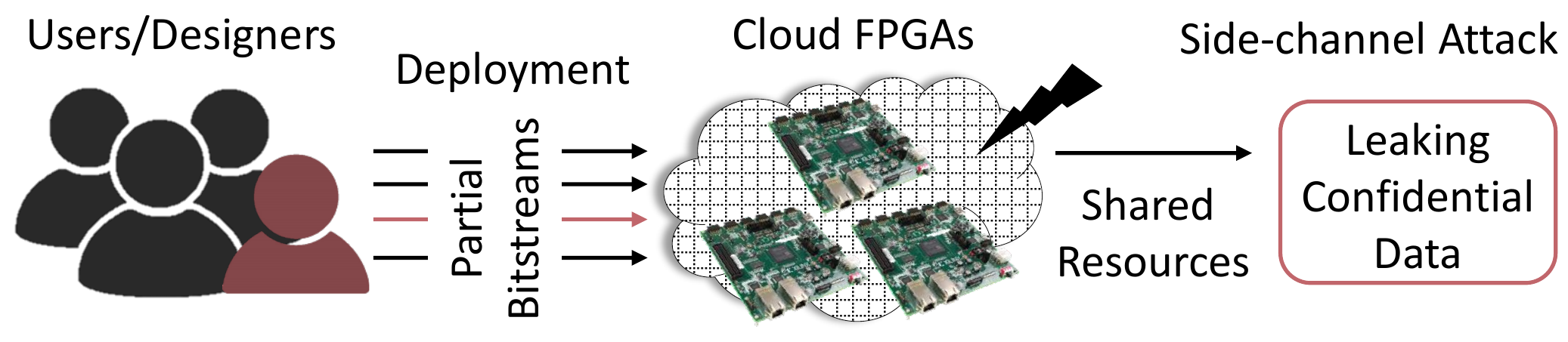}
 \caption{Shared resources in cloud FPGAs leak secret data to malicious users.}
  \vspace{-1em}
 \label{fig:PaaS}
\end{figure}

New security vulnerabilities and attack surfaces arise due to the shared resources among the tenants.  
Various attacks are reported in the literature on remote FPGAs, such as fault injection, covert channel, denial-of-service, and side-channel attacks (SCAs)~\cite{Survey,Survey6}. 
 For example, an attacker can leak critical information of co-located victim tenants via the shared power distribution network (PDN)~\cite{Sensors}, i.e., a power SCA (see Fig.~\ref{fig:PaaS}). 
  In such an attack, an adversary (i.e., malicious tenant) inserts on-chip power sensors, such as ring oscillators (RO)~\cite{zhao_2018} or time to digital converters (TDC)~\cite{cpamap_2020} to record the power trace of a victim tenant and extract secret assets.\footnote{We study a remote side-channel attack scenario where the attacker extracts information from the cloud without physical access to the platform~\cite{zhao_2018}.}
 
 Besides the conventional mitigation techniques against power SCAs, it is imperative to investigate FPGA-assisted defense techniques in this new remote attack surface~\cite{Survey2,Survey3}. The state-of-the-art (SOTA) defenses and their limitations are discussed next and summarized in Table~\ref{tab:comparison}.

\subsection{State-of-the-art and Their Limitations}

 \textbf{Offline Bitstream Scanning:} Pioneer CSPs, e.g., Amazon\textsuperscript{\textregistered}~\cite{amazon}, check the FPGA bitstream and prevent the deployment of combinational loops on the cloud as power measurement sensors~\cite{bitstream2, bitstream3,bitstream4}.  
While these techniques detect some attack circuits, e.g., combinational loops in ROs~\cite{bitstream1,bitstream3}, they cannot identify stealthy attack sensors, e.g., arithmetic-logic units~\cite{gnad2021stealthy}. Also, restricting the hardware design limits the implementation of essential security primitives, e.g., true random number generators. 
 
 \textbf{Run-time Monitoring}  
 detects power fluctuations in shared FPGA platforms (indicating a co-located malicious tenant)~\cite{Monitoring}. Such techniques mitigate active attacks (e.g., fault-injecting) and cannot detect passive remote SCAs~\cite{RO2019}.

 \textbf{Noise Addition:} 
 Defense mechanisms against SCAs decrease the signal-to-noise-ratio to hide the process of the critical assets.
 By introducing noise to the run-time power trace, attackers need more power traces to extract the secret asset (e.g., crypto key). 
 However, these techniques suffer from extremely high area/power overhead (e.g., 100\% area overhead in~\cite{ActiveNoise}), making them impractical in real-world applications.

\begin{table}[!t]
\centering
\caption{Comparison of the state-of-the-art methods against remote power SCAs in cloud FPGAs
}
\label{tab:comparison}
\resizebox{0.48\textwidth}{!}{%
\setlength\tabcolsep{1pt} 
\renewcommand\arraystretch{1.1}
\begin{tabular}{cccc}
\hline
Defense & \begin{tabular}[c]{@{}c@{}} Unrestrained \\ Design\end{tabular} & \begin{tabular}[c]{@{}c@{}}SCAS\\ Prevention\end{tabular}  & \begin{tabular}[c]{@{}c@{}}Low Overhead \\ (Area)\end{tabular} \\ \hline
Offline Bitstream Scanning~\cite{bitstream2}& No & Yes &  $-$\\ \hline
Run-time Monitoring~\cite{Monitoring}& Yes & No &  No ($>$30\%)\\ \hline
Noise Addition~\cite{ActiveNoise} & Yes & No & No (100\%)\\ \hline
Implementation Diversity~\cite{DPR1} & No & Yes & No ($>$2x) \\ \hline
\textbf{Proposed FPGA-Patch} & \textbf{Yes} & \textbf{Yes} & \textbf{Yes} (14\%)\\ \hline
\end{tabular}%
}
\vspace*{-6mm}
\end{table}
 
 \textbf{Implementation Diversity:} Recently, researchers have proposed diversity-based techniques to obfuscate the power trace by exploiting modern FPGA's dynamic partial reconfiguration (DPR) feature~\cite{DPR1,DPR3}. These methods use isofunctional variants swapping to hide the correlation between device power consumption and the target core's intermediate values. However, while diversity-based solutions offer strong defense, they currently have the following limitations that need to be addressed.

 \begin{itemize}[leftmargin=*]
 \item SOTA offer minimal SCA security due to the limited number of deployed variants. 
 Asghar \textit{et al.}~\cite{DPR3} proposed netlist randomization, generating 3 classes of 9 netlist variants, with increasing the attacker's effort only by $\sim$3.5$\times$.
 
 \item The SOTA techniques modify the entire design randomly. Thus, each variant is as large as an entire design bitstream, e.g., in~\cite{DPR1}, 790 MB is required to store $128$ variants. In addition, large bitstream sizes impose large reconfiguration latency and resource utilization overheads. For example, in~\cite{DPR1}, increasing the resistance against power SCAs by two factors caused the resource utilization to double. 
 \item SOTA solutions have been demonstrated via simulations only. The hardware implementation and physical properties of the target platform add technical challenges, such as noise due to process variation in the chip, which cannot be accurately considered in only-simulation results~\cite{DPR4}.

 \end{itemize}

There is a gap in designing lightweight and effective diversity-based obfuscation defenses. Toward that end, we propose \textit{FPGA-Patch}, the first-of-its-kind concept that employs automated program repair methods to generate design variants and thwart power SCA in shared FPGAs. FPGA-Patch is a proactive lightweight defense that cloud FPGA users can employ at the design time prior to bitstream generation. Moreover, FPGA-Patch is generic and can protect any given hardware design. Our novel contributions are as follows.

\subsection{Our Novel Contributions}
\begin{enumerate}[leftmargin=*]
  \item \textbf{Hardware patching for variants generation (Sec.~\ref{Variants}).} FPGA-Patch injects errors into the design to generate faulty netlists, which are then recovered using equivalent checker methods to create diverse power traces at run time. 

 \item 
 \textbf{Eradicating large storage overhead (Sec.~\ref{Bitstream}).} FPGA-Patch adopts a difference-based bitstream generation technique in FPGA design tools to decrease the storage overhead and reconfiguration latency. 

 By limiting the location of modified fault points, FPGA-Patch enables compressed bitstream generation.
 
 \item 
 \textbf{Dynamic FPGA hardware implementation (Sec.~\ref{Runtime}).} FPGA-Patch is implemented on an AMD/Xilinx ZYNQ FPGA using dynamic function exchange (DFX) technology. We evaluate the security of an advanced encryption standard (AES) application core protected by FPGA-Patch against a correlation power analysis (CPA) attack using remote power measurement techniques.
 
\end{enumerate}

\textbf{Key Results.} FPGA-Patch employs 128 variants to thwart the CPA attack, resulting in an increase in the minimum traces to disclosure (MTD) by over 1,000$\times$ and a $2.25\times$ decrease in the CPA value. These improvements significantly reduce the attacker's confidence in key detection. Our experiments show that this enhanced security comes at a minimal area overhead of 14.2\% while preserving the performance of AES.

\section{Background}
\subsection{Remote SCA and Power Sensors (TDC)}
Attackers exploit the shared resource of the PDN in cloud FPGAs by enabling power monitors in the FPGA. To collect power consumption traces, researchers have proposed logical circuits implemented by attackers, such as delay elements like ROs and TDCs. Since ROs are banned from cloud FPGAs by CSPs, TDCs are a valid alternative. As invasive probing techniques are not practical for cloud FPGAs, power monitors are the main method used to collect power consumption traces.

In this work, we adopt TDCs based on the tapped delay lines in~\cite{TDC}, implemented for the Zynq FPGAs, to collect the power traces for offline power analysis in CPA.

\subsection{Logic Equivalence Checking}
Logic equivalence checking (LEC) verifies the functional equivalence of two versions of a design. The LEC tool compares the netlists of the two designs to identify differences and generates a report of faults in the revised logic~\cite{LEC1}. The report can be used to repair any faults found. LEC is commonly used to verify the functionality of a modified or optimized design.

 \section{Threat Model of Proposed FPGA-Patch}

In power SCAs on multi-tenant FPGAs, the attacker is a malicious tenant that exploits the shared PDN~\cite{zhao_2018}. 
Consistent with SOTA work, we assume that the attacker knows the target application running on the victim's partial region. This assumption enables a security assessment of FPGA-Patch in the absence of obscurity. Further, we assume the worst-case scenario for the defender, i.e., when there is no noise added by other tenants' computation in the attacker's observed traces.
  
  Considering crypto cores, the attacker requests the victim to encrypt plain text and observes the produced power traces.
\section{FPGA-Patch Methodology}
\label{Methodology}

 Power SCAs collect power trace samples during multiple executions using power measurement circuitry. These traces are then aligned offline and statistically analyzed to isolate key-correlated parts of each trace. Adding uncertainty to the power profile of the circuit makes trace alignment challenging, reducing the correlation between consumed power and the secret key, thus hampering the SCA. FPGA-Patch tackles this issue by frequently switching between isofunctional variants with diverse power profiles, which enforces heterogeneity in power traces. Fig.~\ref{fig:Methodology} illustrates the integration of FPGA-Patch into the standard FPGA design flow.\footnote{FPGA-Patch is an effective and versatile defense against SCAs targeting the key-dependent execution part of cryptography cores. However, it has limitations, as it only applies to designs where correlation power analysis is valid and does not defend against attacks on machine learning systems, such as side-channel-based model stealing attacks on deep neural networks.}

\subsection{Design of Equivalent Function Variants}
\label{Variants}
FPGA-Patch uses LEC for automatic program repair to generate variants. Fig.~\ref{fig:Methodology}~\Circled{\textbf{A}} shows the standard FPGA design flow, where the hardware's register transfer level (RTL) description is simulated for functional testing, synthesized, and mapped to look up tables (LUT)s. To create design-equivalent heterogeneous variants, FPGA-Patch selects candidate nets in the target design netlist for fault injection (See Fig.~\ref{fig:Methodology}~\Circled{\textbf{B}}).

\subsubsection{Net selection}
We consider two net selection methods, discussed below. The number of selected nets is a parameter to evaluate the diversity of variants.

\textbf{Random Selection.} For our initial exploration, we consider random net selection. However, some selected nets could not change the run-time power characteristics effectively. 

\begin{figure}[!t]
 \centering
 \includegraphics[width=\linewidth]{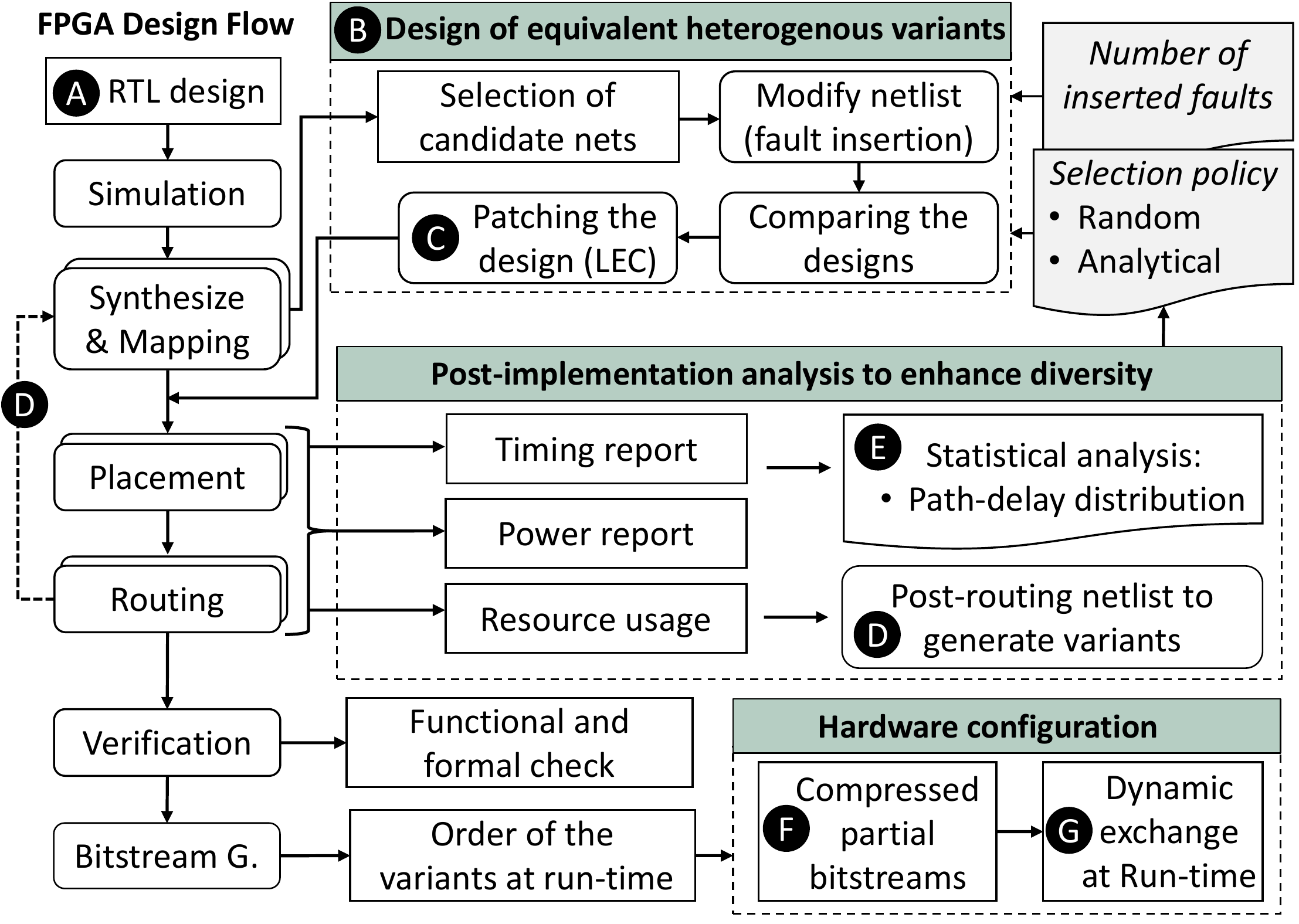}
 \caption{Proposed FPGA-Patch methodology for mitigating power SCAs. }
  \vspace{-1.5em}
 \label{fig:Methodology}
\end{figure}

\textbf{Critical Net Selection.} We modify FPGA-Patch to select ``effective'' candidate nets, increasing the diversity of variants.
Selecting nets on the critical timing path, maximizes the probability of run-time power obfuscation, see Fig.~\ref{fig:Methodology}~\Circled{\textbf{E}}.

\subsubsection{Fault Insertion}
Selected nets are connected to the ``0'' and ``1'' logical ports, resembling stuck-at-0 (SA0) and stuck-at-1 (SA1) faults, respectively. After fault injection, the design tool optimizes the netlist and removes unnecessary logic gates, unused ports, and unconnected wires. These manipulations are intended to alter design functionality.

\subsubsection{Design-Level for Fault Insertion}
We explore the difference between injecting the faults at the post-synthesis netlist versus the post-routing netlist. The proposed flow is the same in both cases. More details are provided in Sec.~\ref{sec:fault_location}.

\subsubsection{Patching the Faulty Netlists}
\label{patch}
Each faulty netlist should be patched for missing logic using the LEC methods, as shown in Fig.~\ref{fig:Methodology}~\Circled{\textbf{C}}. 
To recover the target design's functionality, each faulty netlist is compared with the netlist of the original design.
Formal equivalence checking can catch any formal errors in combinational logic, known as Nonequivalents (NEQs).
To repair the faulty netlist, the LEC traces back the NEQs to find the unmapped or incorrectly mapped nets. After detecting all NEQs, the LEC adds the logic required to recover each missing state to the design steadily until all NEQs are resolved. 
 
\subsection{More on Fault Injection}
\label{sec:fault_location}

\subsubsection{Post-Synthesis} 
\begin{figure}[!t]
 \centering
 \includegraphics[width=\linewidth]{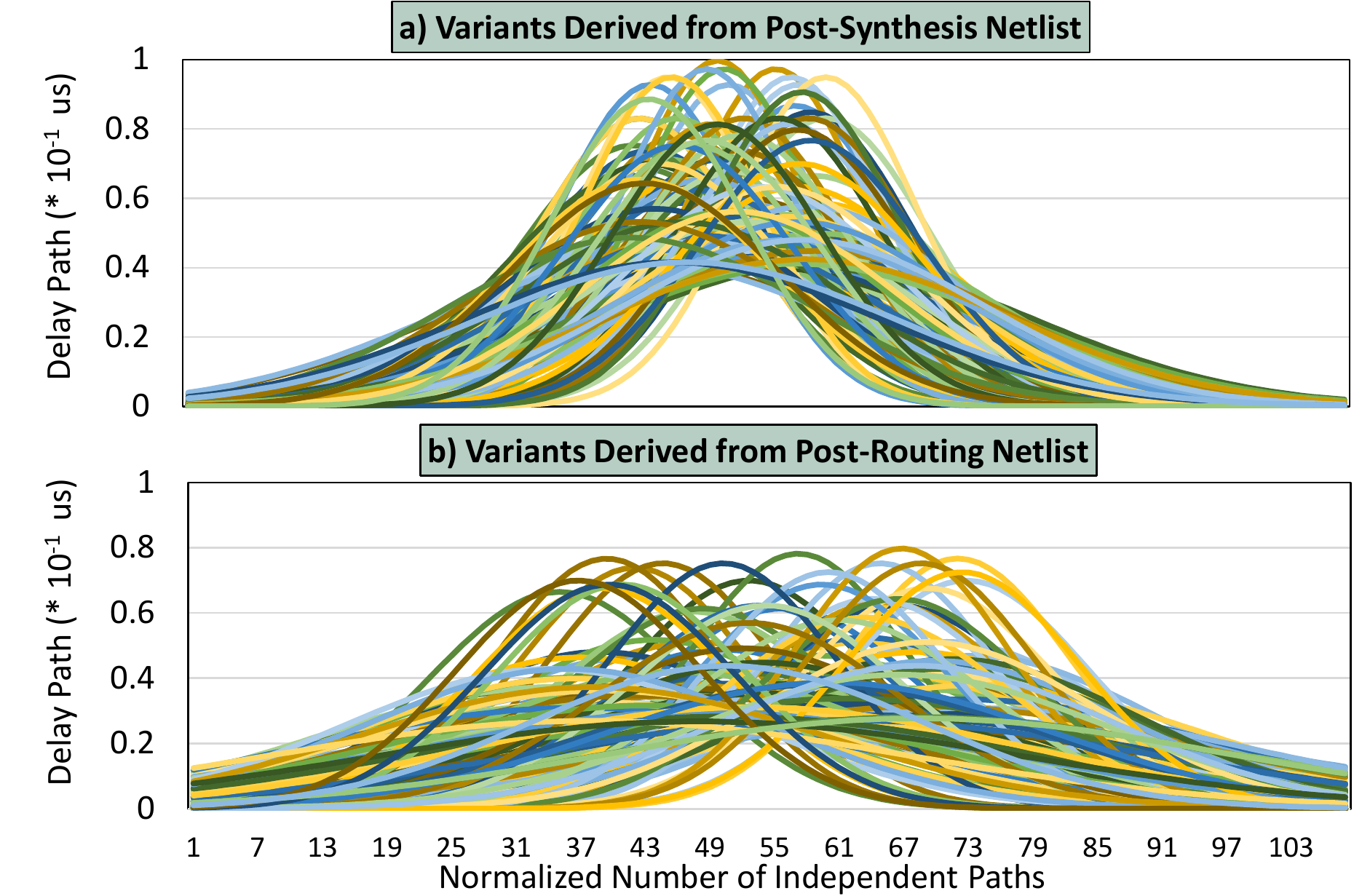}
 \caption{Timing comparison of 128 variants generated by FPGA-Patch regarding the delay of independent paths in one clock cycle. The lines represent the best Gaussian fit of the estimated path delays of routed design.}
 \vspace{-1.5em}
 \label{fig:DelayPath}
\end{figure}

For the initial exploration, we inject the faults into the post-synthesis netlist and study the resource utilization of the different variants, as shown in Fig.~\ref{fig:Methodology}~\Circled{\textbf{D}}. Fig.~\ref{fig:DelayPath}(a) indicates the delay distribution between 128 generated variants, considering AES as a target application. As can be observed, the diversity between the variants is limited. Our investigation indicates that the altered nets were different bits of the same data bus. Therefore, they all had a similar effect on the hardware implementation. Further, passing the recovered netlists through the rest of the FPGA design flow steps, i.e., routing, etc., applies logic optimization methods, eliminating redundant paths and leading to lower diversity.

\subsubsection{Post-Routing}
Next, we investigate the effect of injecting the faults into the post-routing netlist, see Fig.~\ref{fig:Methodology}~\Circled{\textbf{D}}, and present the path delay distribution for the 128 variants in Fig.~\ref{fig:DelayPath}~(b). 

We can infer that the distribution of maximum delay of independent paths for post-routing is more diverse compared to post-synthesis. This diversity leads to variable dynamic power when switching between the variants at run time and hiding the secret key's power trace. \textit{Therefore, we recommend that FPGA-Patch gets adopted post-routing.}

\subsection{Hardware Configuration File}
\label{Bitstream}

When the differences between two versions of the  hardware are minimal, we can use difference-based bitstream generation to generate configuration files; see Fig.~\ref{fig:Methodology}~\Circled{\textbf{F}}. 
Difference-based bitstream generation can be applied to the designs in which the changes among them are localized.

This compressed partial bitstreams program only the difference between two given variants. Switching the configuration of a module from one implementation to another is fast because the bitstream differences are smaller than the entire partial reconfiguration region (PRR) bitstream. 

Initially, on device power-up, a complete bitstream must be loaded into the device prior to any partial bitstreams. Therefore, a full bitstream configuring all PRRs to initial configurations needs to be loaded. After the first reconfiguration, a partial bitstream will be loaded to each PRR. During the reconfiguration process, the states of the flip-flops are preserved, and the fixed parts of the design remain fully operational. This ensures that reconfiguration does not affect the performance of the target application and FPGA-patch does not impose performance overhead, in terms of the throughput of the encryption core.

\subsection{Dynamic Exchange at run-time}
\label{Runtime}

FPGA-Patch obfuscates confidential applications' dynamic power trace by changing the underlying hardware dynamically. A difference-based method is deployed to generate the configuration file of the variants, where the distance vector of two consecutive designs is stored in memory. Therefore, the order of deployed bitstreams in a specific PRR is determined at design time. We divide the 128 diverse designs into 8 categories, and each category is assigned to one PRR. Hence, we can avoid stalls in the execution of the encryption algorithm during the reconfiguration via cycling around the PRRs and inside the categories per each encryption. The partial bitstreams in PRR use a fixed order of configurations independently (C1 to C2 to C4,..., to C8).

The target module activates one PRR at a time while the other PRRs undergo reconfiguration with the next variant. To ensure the required time to finish the next module reconfiguration and preserve the throughput, the active time of each variant should be less than MTD for unprotected design. The DFX module in Xilinx/AMD FPGAs automatically programs the next PRR to be used in the reconfiguration process, as illustrated in Fig.~\ref{fig:Methodology}~\Circled{\textbf{G}}.

\section{Experimental Investigations}

\begin{figure}[!t]
 \centering
 \includegraphics[width=\linewidth]{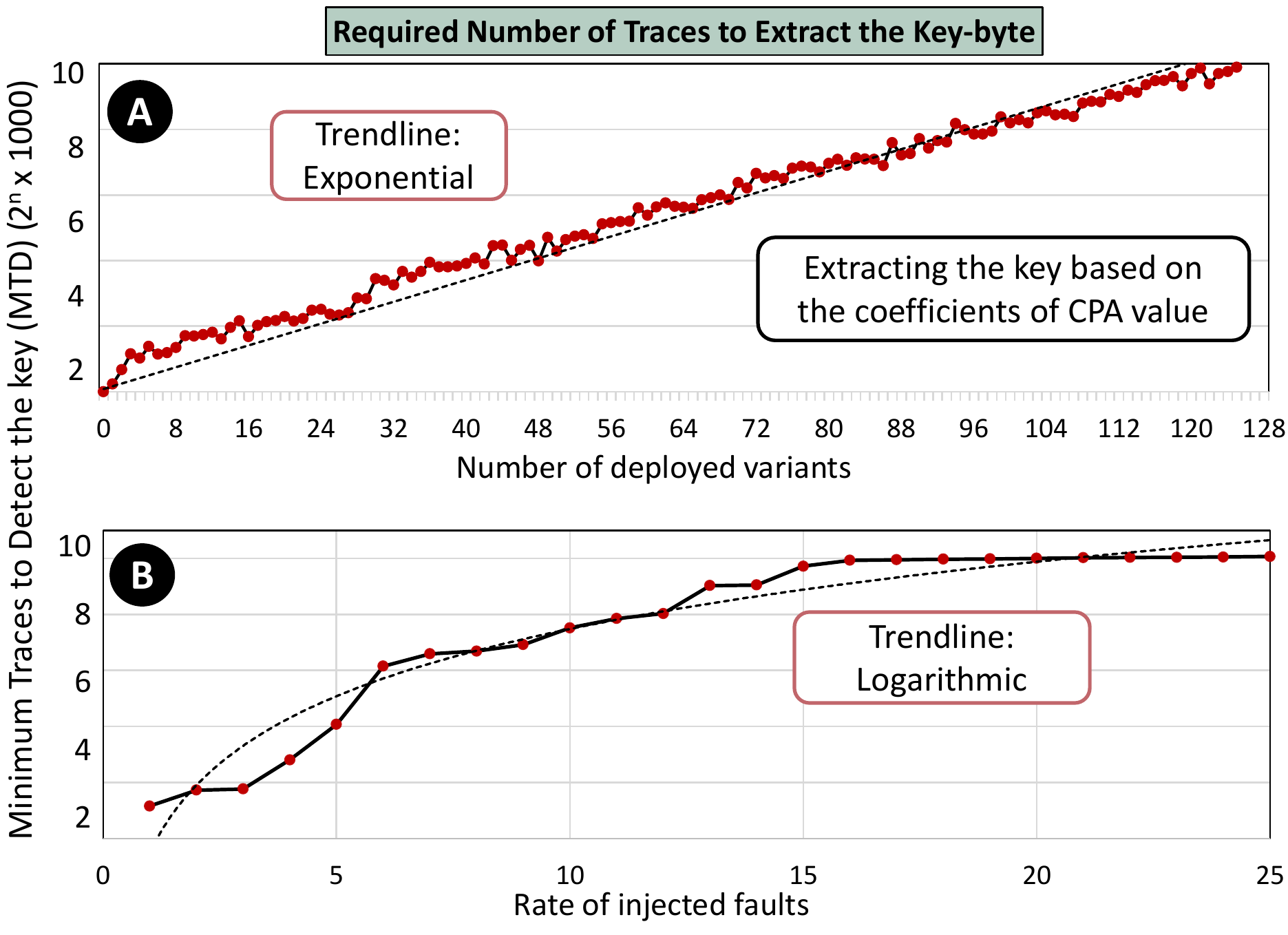}
  \vspace*{-8mm}
 \caption{Analysis of the number of deployed variants on attacker's effort in terms of MTD, when the injected fault rate is 10\%.}
  \vspace{-1.5em}
 \label{fig:Traces}
\end{figure}

\begin{figure*}[!t]
 \centering
 \includegraphics[width=\textwidth]{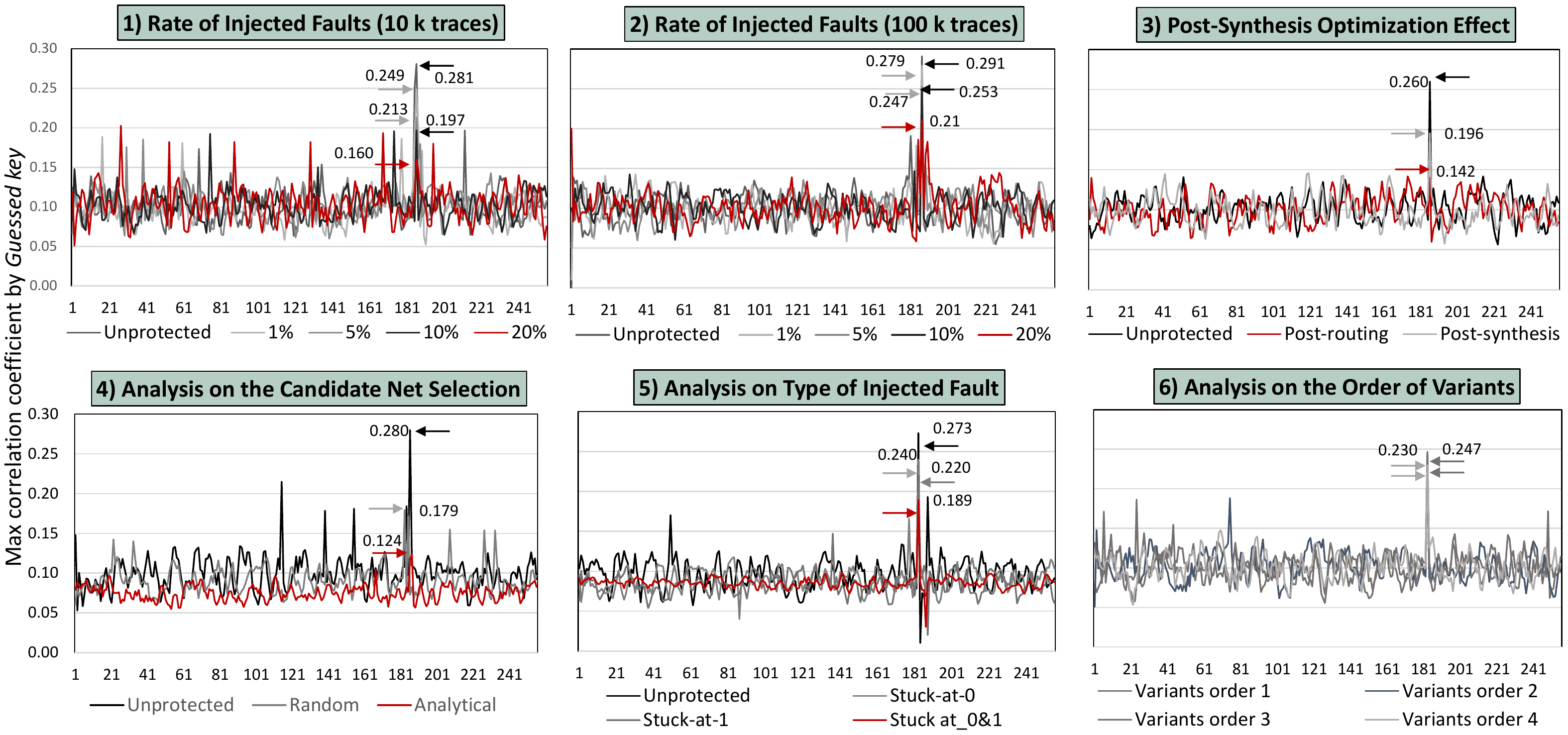}
  \vspace{-2em}
 \caption{Max correlation coefficient analysis by guessed key, with correct key-byte of 185 (0xB9). The red line highlights the best-obfuscated results.}
  \vspace{-1em}
 \label{fig:Corr}
\end{figure*}

To evaluate the effectiveness of the proposed FPGA-Patch, we analyze it in a real-setup implementation on the FPGA. First, we describe the experimental setup, then we describe the security improvements and compare them with the SOTA.
\subsection{Experimental Setup and Tool Flow}
\textbf{Target Hardware.}  
Following SOTA works~\cite{DPR1,DPR2}, we use a serial AES-128 encryption core, employed by a security application encrypting a test dataset. 
The S-box operation which is the focus of design diversity is implemented using Gallois field implementation taken from opencores.org~\cite{opencores}.

After the first round (SubBytes), the state is a function of input data and the encryption key. 
Hence, the SubByte is considered as the operation-to-be-protected by FPGA-Patch.

\textbf{FPGA Implementation.} The AES core is implemented on the ZedBoard Zynq-7000 ARM/FPGA SoC development board, using Vivado Design Suite'21, requiring 256 LUTs and 260 registers. Enabling the I/O interfaces requires additional standard IPs and AXI peripherals, which makes the static region require 422 CLBs in Zedboard (1277 LUTs, 2024 registers).

For \textbf{Fault Injection}, we adopt the engineering change orders (ECOs) flow. ECOs are modifications to the design netlist to implement the changes with minimal impact on the original design. 
Modern FPGA design methods, i.e., Vivado, provide an ECO flow, which allows modification of a design checkpoint, implements the changes, runs reports on the changed netlist, and generates the final programming files. FPGA-Patch leverages ECO to select and modify the netlist to ensure the semantics of the design netlist after fault injection.

\textbf{Variants.} To study the security of FPGA-Patch, 128 variants are generated for each experiment, which are divided into 8 categories for exchanging at run time. Following the design constraints, operating frequency of AES is set to 10 MHz.

\textbf{Dynamic Exchange.} The FPGA processor hosts a Linux OS to control the I/O interface, reconfiguration application, and the AES IP core data transfer. 
To enable high-speed reconfiguration of PRR, we adopt the internal configuration access port implemented for DFX procedure in Vivado.

 The \textbf{Order of Variants} 
 generated by FPGA-Patch does not impact its security performance. The correlation equation considers all collected traces, ensuring that the distribution of variant usage is fixed. Even if an attacker obtains and re-organizes all the traces, the security remains intact as long as this distribution is maintained.

\textbf{Attack Measurement.} We collect the power traces of FPGA-Patch using a TDC~\cite{TDC} with 2 channels, running in an independent PRR. A single encryption run corresponds to one measured trace. The output is periodically sent to a host workstation application for further analysis via the CPA technique.
As an evaluation metric, we perform a CPA~\cite{CPA}, with the aforementioned Hamming Distance power model targeting two consecutive S-box outputs during the first round. Then, we compute the CPA value for each key-byte guess and for each power trace based on the number of traces.

\begin{table*}[!t]
\renewcommand*{\arraystretch}{1.2}
\centering
\caption{Overhead Comparison of the Injected Fault Rate in FPGA-Patch Methodology}
\label{tab:Overhead}

\resizebox{\textwidth}{!}{%
\begin{tabular}{|c|c|c|c|c|c|c|c|c|c|c|c|c|c|c|c|c|c|c|c|c|c|c|c|c|c|}
\hline
\textbf{Fault Rate (\%)} & 1    & 2    & 3   & 4   & 5   & 6    & 7    & 8    & 9    & 10   & 11   & 12   & 13   & 14   & 15   & 16   & 17   & 18   & 19   & 20   & 21   & 22   & 23   & 24   & 25   \\ \hline
\textbf{Area (LUT) (\%)} & 3.3  & 4.3  & 6.6 & 7.3 & 8.1 & 12.1 & 12.5 & 13   & 13.2 & 14.2 & 14.7 & 15.2 & 18.5 & 22.4 & 23.2 & 25.9 & 31.5 & 32   & 35.7 & 37.5 & 47.4 & 50.9 & 51.5 & 52.3 & 54   \\ \hline
\textbf{Power (\%)}      & 0.6  & 0.8  & 1.2 & 1.3 & 2.1 & 5.2  & 6.7  & 6.9  & 7    & 7.3  & 7.5  & 7.6  & 8.4  & 9.1  & 9.4  & 10   & 10.2 & 10.3 & 10.5 & 12.2 & 15.1 & 17.4 & 17.7 & 17.8 & 18   \\ \hline
\textbf{Path Delay (\%)} & 0.18 & 0.19 & 0.2 & 0.2 & 0.2 & 0.21 & 0.22 & 0.22 & 0.22 & 0.24 & 0.24 & 0.26 & 0.27 & 0.27 & 0.29 & 0.29 & 0.3  & 0.32 & 0.33 & 0.33 & 0.34 & 0.38 & 0.39 & 0.39 & 0.39 \\ \hline
\end{tabular}
}
\vspace*{-5mm}
\end{table*}

\subsection{Security Analysis}
We analyze the security of FPGA-Patch against CPA in terms of (i)~MTD (see Fig.~\ref{fig:Traces}) and (ii)~the correlation coefficient of all key-bytes compared to the correct key-byte (see Fig.~\ref{fig:Corr}).

\textbf{Attacker's Effort.} The attacker attempts to find the correct key-byte by correlating power samples with the hypothetical power model of each possible byte value for the first byte of the key. The values that are $1.5\times$ higher than the rest are considered the correct key. The MTD of AES with respect to the \#variants deployed at runtime is shown in Fig.~\ref{fig:Corr}~\Circled{\textbf{A}}. Assuming that the correct key-byte in a CPA against AES can be detected in $\approx 1000$ traces, deploying 1 to 128 variants in FPGA-Patch results in an exponential growth of the MTD for various numbers of traces. By applying 128 variants generated with 10\% injected faults to the post-routing netlist while selecting nets analytically (as explained in Sec.~\ref{overhead}), the MTD has increased by at least three orders of magnitude.

 The diversity of the variants depends on the number of injected faults. Our analysis reported in Fig.~\ref{fig:Corr}~\Circled{\textbf{B}}, shows that injecting more than 10\% of faults increases the MTD only by 4x, where the threshold of circuit recovery by LEC, has been observed to be $\sim$25\%. Therefore, the effective injected fault rate for random net selection policy is identified to be 10\%.

\textbf{Attacker's Confidence.} The maximum correlation coefficient of each key-byte guess and assumed power model is shown in Fig.~\ref{fig:Corr}. Details of the results are explained below.

 \textbf{Effect of the Injected Fault Rate.} The rate of injected faults is defined as the number of candidate nets to assign the fault divided by the total number of nets in the DPR module (SubByte).
The CPA value for 128 variants deployed at run time is compared for different fault injection rates. The comparison indicates that increasing the number of randomly injected faults decreases the CPA value by $1.75\times$ when the most repairable nodes are faulty.

\textbf{Increasing \#Traces.} Attacker's confidence in extracting the secret key-byte, is defined by CPA value for a specific number of traces to analyze. As shown in Fig.~\ref{fig:Corr} 1 and 2, collecting and analyzing 10x more power traces, eliminates the false positive in results and increases CPA value by 0.27x. 

 \textbf{Post-Synthesis Fault Injection vs. Post-Routing.} The experiments on the adopted netlist from post-synthesis and post-routing, show that applying faults to the routed design is $1.83\times$ more effective than only synthesized netlist, in terms of decreasing the CPA. The optimization in re-synthesis stage, eliminated the effect of patch logic on power obfuscation.

\textbf{Effect of Node Selection.} The candidate nets to inject the faults are selected randomly and analytically (based on the critical path delay order). Selection of the nets randomly decreases the CPA value for $1.56\times$, while selecting nodes based on the timing reports decreases the CPA peak for $2.25\times$.

\textbf{Effect of Fault Type.} To study the impact of the type of injected faults, the CPA value for SA0 and SA1 and both, are shown in Fig.~\ref{fig:Corr}.5. It shows the value of CPA for both SA0 and SA1 are in similar trends, while when the both are deployed, the peak of CPA value decreases by $1.44\times$.

\vspace*{-2mm}
\subsection{Performance Overhead}
\label{overhead}
We have explored a fault rate in the range from 1\% to 25\%, which is the maximum repairable threshold for LEC. The trade-off between security and overhead shows that the area overhead, from 10\% to 25\% increases by $2.8\times$, while increasing MTD by only $4\times$. Therefore $10\%$ is selected as an effective fault rate in our experiments.

\textbf{Area Overhead.} Exploiting the optimized netlist and adding logic to repair the faults increases the number of logic gates and the used area. As shown in Table~\ref{tab:Overhead}, the area overhead increases with the number of faults injected into the design. In our security analysis, considering the effective fault rate as 10\%, the area overhead is 14.2\% in FPGA-Patch.

 \textbf{Delay Overhead.} 
 FPGA-Patch affects delay only if the candidate nets affect the critical paths. The effect of FPGA-Patch on the average path delay is minimal. Considering 10\% fault rate, the path delay overhead is 0.24\%.

 \textbf{Power Overhead.} The overhead of dynamic power in FPGA-Patch includes the reconfiguration controller and switching activity of the hardware. 
 Moreover, our analysis indicates that when candidate nets are selected based on the timing report, the increase in dynamic power is 12\% more than when selected randomly. 
 
 \textbf{Memory Storage Size.} A full bitstream of AES is $\sim$460 kB, and each partial bitstream depending on the injected fault rate is $\sim$ 10 to 50 kB. The total storage size required for full bitstream and 128 variants is 1.8 to 7 MB, whereas, in module-based bitstream generation, the required storage is $\sim$60 MB. 
 
 \textbf{Performance overhead.} FPGA-Patch's small variant bitstream size enables fast reconfiguration. Further, the cyclic selection of PRRs guarantees an active configuration for encryption, preserving AES's throughput and performance.

 \vspace{-0.5 em}
\subsection{Comparison with State-of-the-Art}

Recently, design diversity enabled by DPR, has been studied as a defense method against SCA.
In this section, we compare our findings with three approaches~\cite{DPR1,DPR3,moving} that advocate for the dynamic exchange of hardware variants, summarized in Table.~\ref{tab:Related}.

In~\cite{DPR1}, the variants are diversified in placement and routing strategies. This work deploys 128 AES variants to exchange at runtime, and the gained security in terms of confidence of attacker, in detecting the key is limited to $\sim$2.2$\times$.
Similarly, Asghar et al. in~\cite{DPR3} have proposed netlist randomization techniques that randomly enable delay elements in the design. By developing 9 netlist variants from three classes of diversity, they show the peak of CPA value has decreased by $\sim$3.54$\times$.
Moreover,~\cite{DPR3} employs long chains of delay elements and dummy modules, which increases the hardware resource overhead by $\sim$5$\times$. 
Furthermore, in ~\cite{moving}, several reconfiguration regions are reserved in the FPGA, and the design is moved among these regions in each reconfiguration, in addition to noise generators to further obfuscate the power traces. While in ~\cite{moving} the authors show MTD is increased by $\sim$95$\times$, the security of this technique is bounded by the process variation on the FPGA SoC. 

In contrast, the design size in FPGA-Patch exhibits negotiable changes, $<$14.2\%, for resource overhead (due to adding patch logic) to AES, increasing the attacker's effort to extract the correct key-byte by $>3\times$.

\begin{table}[!t] 
\centering
\caption{Comparison with related works }
\label{tab:Related}
\resizebox{0.48\textwidth}{!}{%
\setlength\tabcolsep{6pt} 
\renewcommand\arraystretch{1.1}
\begin{tabular}{ccccc}
\hline
              Comparison Category     & \cite{DPR1} & \cite{DPR3} & \cite{moving}& FPGA-Patch \\ \hline
Increased MTD      &  ~2.2$\times$    & ~3.54$\times$   & ~95$\times$  &   ~1000$\times$        \\ \hline
Number of variants &    128  &  9    & 16   &    128       \\ \hline
Throughput overhead &    NA  & $\sim$5$\times$   &  0  &   0        \\ \hline
\end{tabular}
}
\vspace*{-6mm}
\end{table}

\vspace*{-2mm}
\section{Conclusion}
CSPs are highly concerned about securing shared platform applications against remote SCA. To address this concern, we propose \textit{FPGA-Patch}, a novel defense that utilizes program repair algorithms to prevent SCAs on cloud FPGAs. By generating isofunctional variants of the target hardware and dynamically exchanging them at runtime, we can obfuscate power traces and hinder SCA. The variant generation policy of FPGA-Patch allows for bitstream compression and minimizes dynamic exchange costs, making it a lightweight and effective diversity-based obfuscation defense against remote SCAs. Overall, FPGA-Patch provides a promising solution for CSPs seeking to enhance the security of their shared platforms.
\vspace*{-3mm}

\bibliographystyle{IEEEtran}
\bibliography{main}

\end{document}